%% file: main.tex
\documentclass[runningheads]{llncs}
\usepackage[T1]{fontenc}
\usepackage{booktabs}
\usepackage{csvsimple}
\usepackage[dvipsnames]{xcolor}
\definecolor{lightorange}{RGB}{255,230,204} 
\input{preamble}

\begin{document}

\title{Learning Segmentation from Radiology Reports}
\titlerunning{Learning Segmentation from Radiology Reports}

\author{
Pedro R. A. S. Bassi\inst{1,2,3} \and
Wenxuan Li\inst{1} \and 
Jieneng Chen\inst{1} \and \\
Zheren Zhu\inst{4,5} \and
Tianyu Lin\inst{1} \and
Sergio Decherchi\inst{3} \and 
Andrea Cavalli\textsuperscript{2,3,6} \and \\
Kang Wang\inst{4} \and
Yang Yang\inst{4} \and
Alan L. Yuille\inst{1} \and
Zongwei Zhou\thanks{Correspondence to: Zongwei Zhou (\href{mailto:zzhou82@jh.edu} {\texttt{zzhou82@jh.edu}})}\inst{1}
}

\authorrunning{P. R. A. S. Bassi et al.}
\institute{
John Hopkins University \and
University of Bologna \and
Italian Institute of Technology \and
University of California, San Francisco \and
University of California, Berkeley \and
École Polytechnique Fédérale de Lausanne
}

\maketitle              

\begin{abstract}
Tumor segmentation in CT scans is key for diagnosis, surgery, and prognosis, yet segmentation masks are scarce because their creation requires time and expertise. Public abdominal CT datasets have from dozens to a couple thousand tumor masks, but hospitals have hundreds of thousands of tumor CTs with radiology reports. Thus, leveraging reports to improve segmentation is key for scaling. In this paper, we propose a report-supervision loss (\method) that converts radiology reports into voxel-wise supervision for tumor segmentation AI. We created a dataset with 6,718 CT-Report pairs (from the UCSF Hospital), and merged it with public CT-Mask datasets (from AbdomenAtlas 2.0). We used our \method\ to train with these masks and reports, and strongly improved tumor segmentation in internal and external validation---F1 Score increased by up to 16\% with respect to training with masks only. By leveraging readily available radiology reports to supplement scarce segmentation masks, \method\ strongly improves AI performance both when very few training masks are available (e.g., 50), and when many masks were available (e.g., 1.7K). Project: \href{https://github.com/MrGiovanni/R-Super}{https://github.com/MrGiovanni/R-Super}

\end{abstract}

\section{Introduction}\label{sec:introduction}

Tumor segmentation in CT scans is key for diagnosis, surgery, and radiotherapy planning. However, creating voxel-wise segmentation masks is time-consuming and expensive \cite{wasserthal2023totalsegmentator,FLARE23-ma2024automaticorganpancancersegmentation,chou2024acquiring,chou2024embracing,li2025pants}
. Thus, most public CT datasets have less than 100 tumor masks, few exceed 1,000, and none exceeds 2,000 masks per tumor type \cite{bilic2019liver,heller2019kits19,antonelli2021medical}. Creating segmentation masks is not part of a radiologist's everyday job, but creating radiology reports (texts explaining the findings in a CT) is. Thus, hospitals have hundreds of thousands of CT-Report pairs, and a recent public dataset has more than 50,000 CT-Report pairs (CT-RATE \cite{hamamci2024ct2rep}).
Furthermore, the information inside reports is valuable for segmentation: tumor count, location (organ / organ sub-segment), and sizes. Radiology reports seem to be a large-scale and valuable source of information, but \textit{can reports improve tumor segmentation?}
To answer this, studies extracted classification labels from reports (e.g., tumor present / absent) and used them in multi-task learning for segmentation and classification \cite{zhang20213d}. Studies also used reports in contrastive language-image pre-training (CLIP) \cite{blankemeier2024merlin}. These methods use reports to optimize auxiliary tasks (classification / CLIP), not to optimize segmentation itself. However, the benefit of auxiliary tasks to segmentation is indirect and sometimes minor (Fig.~\ref{fig:all_plots}).

We hypothesize that radiology reports can be used to \textit{directly supervise} tumor segmentation. We propose \method\ (\textbf{R}eport-\textbf{Super}vision), a training paradigm that introduces two novel loss functions—Volume Loss (\S\ref{sec:volume_loss}) and Ball Loss (\S\ref{sec:ball_loss})—to enforce alignment between segmentation outputs and tumor attributes extracted from reports (count, size, and location). This information is parsed by a large language model (LLama 3.1 \cite{dubey2024llama}) using radiologist-designed prompts. \method\ applies to any model architecture and, unlike prior methods that rely on auxiliary tasks, \method\ supervises segmentation directly.

To train \method, we built a massive dataset sourced from the University of California San Francisco (UCSF) Hospital and nearby institutions (California, USA). Like most hospitals, UCSF could not provide segmentation masks, only reports. However, \privateOneTrain\ has more tumor CTs (4,967) than any public abdominal CT dataset \cite{jaus2023towards,ji2022amos,li2024abdomenatlas,li2024well,qu2023annotating}. 
We trained \method\ using the \privateOneTrain\ reports, in addition to few (50), medium (344) or many (1.7K) tumor segmentation masks (sourced from AbdomenAtlas 2.0 \cite{bassi2025radgpt}). After training, we conducted both internal validation (\privateOneTest), and external validation on a hospital unseen during training. 
\method\ strongly improved segmentation and surpassed the 5 state-of-the-art methods in Tab.~\ref{tab:sota_comparison}. Our main contributions are:

\begin{enumerate}
    \item We created loss functions that optimize segmented tumors to be consistent with radiology reports. To the best of our knowledge, our \method\ losses are the first to use radiology reports to directly optimize segmentation in CT.
    \item \method\ uses reports to achieve better tumor segmentation with \textit{few} (50), \textit{medium} (344), and \textit{many} (1.7K) segmentation masks in training (up to +16\%, +9.2\%, and +4.3\% F1-Score, respectively, Fig.~\ref{fig:all_plots}).
    \item \method\ improved tumor segmentation both in the hospital that provided the radiology report for training (up to +16\% F1-Score, Fig.~\ref{fig:all_plots}), and on an unseen hospital (up to +9.2\% F1-Score, Fig.~\ref{fig:all_plots}).
\end{enumerate}

\begin{table}[t]
\centering
\scriptsize
\caption{
\textbf{Comparison with related work.} We compare \method\ with five state-of-the-art methods. \textbf{(1)} Standard segmentation uses only CTs with manual masks. \textbf{(2)} Models Genesis applies self-supervision on all CTs but ignores reports. \textbf{(3)} CLIP-Like and \textbf{(4)} Multi-task learning use reports for auxiliary tasks (e.g., classification), not for direct segmentation. \textbf{(5)} Report-guided pseudo-labels reduce false positives but miss false negatives and ignore tumor size. \ul{In contrast}, \method\ uses detailed report information (tumor count, size, location) to directly supervise segmentation and penalize both false positives and false negatives.
}

\label{tab:sota_comparison}
\begin{threeparttable}
\begin{tabular*}{\linewidth}{@{\extracolsep{\fill}}lccccc@{}}
\toprule
training paradigm   & learns from & learns w/ & uses detailed       & reports optimize      & reports penalize \\
                            & CT w/o mask &  reports & report info.$^\dagger$       & segmentation directly       & FP/FN directly            \\
\midrule
Segmentation \cite{isensee2021nnu,gao2022data}      & \xmark      & \xmark  & \xmark                & \xmark             & \xmark           \\
Models Gen. \cite{zhou2021models,zhou2019models}     & \cmark      & \xmark  & \xmark                & \xmark             & \xmark           \\
Multi-task Learn. \cite{chen2019lesion}    & \cmark      & \cmark  & \xmark                & \xmark             & \xmark           \\
RG pseudo-labels \cite{bosma2023semisupervised} & \cmark      & \cmark  & \xmark                & \cmark             & \xmark           \\
CLIP-Like \cite{blankemeier2024merlin} & \cmark & \cmark & \cmark                & \xmark             & \xmark  \\
\midrule
\textbf{\method\ (ours)}    & \cmark      & \cmark  & \cmark                & \cmark             & \cmark           \\
\bottomrule
\end{tabular*}

\begin{tablenotes}[flushleft]
  \item[$\dagger$] AI learns from tumor count, sizes, and locations (organ/organ sub‑segment) in reports.
\end{tablenotes}

\end{threeparttable}
\end{table}

\section{\method}

\newloss\ trains AI to segment tumors that are consistent with the tumor information in radiology reports. \textbf{First}, it extracts (and stores) this information---tumor count, locations (organ/organ sub-segment), and diameters---from reports using a zero-shot LLM, Llama 3.1 70B AWQ. It uses radiologist-designed prompts and has 96\% accuracy extracting tumor presence \& location \cite{bassi2025radgpt}.
\textbf{Second}, we train the tumor segmenter with the available (possibly few) CT-Mask pairs, using traditional segmentation losses (cross-entropy \& DSC). \textbf{Third,} we fine-tune it with both CT-Mask pairs and CT-Report pairs. For the CT-Mask pairs, we use the traditional segmentation losses; for CT-Report pairs, we use two novel losses: \textit{Volume Loss} (\S\ref{sec:volume_loss}) and \textit{Ball Loss} (\S\ref{sec:ball_loss}). The Volume Loss is applied to an intermediate layer of the segmenter (deep supervision), and it constricts segmented tumors to match the tumor locations and overall tumor volumes extracted from reports. The Ball Loss is stricter, and is applied to the last layer of the segmenter. It optimizes each segmented tumor individually, making them match the tumor count, locations, volumes and diameters in the report. \textit{Overall, \newloss\ transforms text-wise report supervision into voxel-wise supervision.}

\subsection{Volume Loss}
\label{sec:volume_loss}

\begin{figure}[t]
    \centering
    \includegraphics[width=1\linewidth]{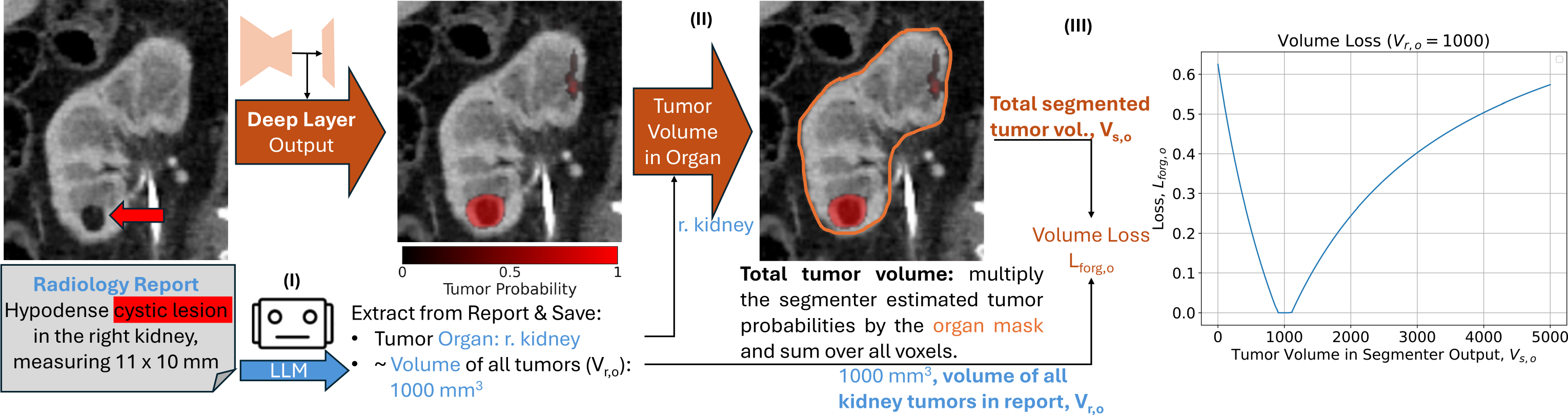}
    \caption{
    \textbf{The Volume Loss (deep supervision) aligns segmented tumors with volumes and locations from reports.} It consists of three steps.
    \textbf{(I)} A large language model (LLM) extracts tumor location (organ/sub-segment) and diameters from the report. Using ellipsoid approximations, we estimate the total tumor volume in each organ $o$ as $V_{r,o}$. 
    \textbf{(II)} From a deep layer of the segmenter, we compute tumor probabilities via a 1$\times$1$\times$1 convolution and sigmoid/softmax. We sum the probabilities inside the organ (using pre-saved organ masks) to estimate the segmented tumor volume, $V_{s,o}$. 
    \textbf{(III)} The Volume Loss penalizes discrepancies between $V_{s,o}$ and $V_{r,o}$. As shown in the right panel (for $V_{r,o} = 1000$), the loss is zero when $V_{s,o}$ is close to $V_{r,o}$, allowing for some tolerance due to uncertainty in report-based volume estimates.
    }
    \label{fig:volume_loss}
\end{figure}

The Volume Loss (Fig.~\ref{fig:volume_loss}) enforces consistency between the total tumor volume in the segmentation and in reports. Instead of volumes, reports usually inform tumor diameters, which we use to estimate volumes. Reports provide one to three diameters per tumor. One diameter is common for small, rounder tumors, and two perpendicular diameters are the World Health Organization standard \cite{miller1981reporting}. With one diameter informed ($d_1$), we estimate tumor volume as a ball volume: $d_1^3 \pi/ 6$. With three diameters ($d_1$, $d_2$, $d_3$), we use the ellipsoid volume: $d_1d_2d_z\pi/ 6$. With two diameters ($d_1$, $d_2$), we estimate the third as $d_3=(d_1+d_2)/2$, and use the ellipsoid volume. For each organ or organ sub-segment ($o$), we sum the volumes of all tumors in the report, giving $V_{r,o}$.
After estimating $V_{r,o}$ from the report, Eq.~\ref{eq:vol_calculation} estimates $V_{s,o}$, the total volume of all tumors the AI \textit{segmented} in $o$. Eq.~\ref{eq:vol_calculation} uses the tumor segmentation, $\bm{T}=[t_{h,w,l}]$\footnote{The volume loss is used as deep supervision. Thus, $\bm{T}$ is created from the output of a deep convolutional layer (e.g., MedFormer decoder layer 2), after a 1x1x1 conv. to reduce channels, sigmoid activation, and linear interpolation to the CT spacing.}; pre-saved, binary organ segmentation masks, $\bm{O}=[o_{h,w,l}]$\footnote{Many accurate public organ segmentation models exist \cite{bassi2024touchstone,wasserthal2023totalsegmentator}. We used an nnU-Net \cite{isensee2021nnu} trained on AbdomenAtlas 2.0 \cite{bassi2025radgpt}, which we made public. It segments 39 organs / anatomical structures, including the pancreas sub-segments (head, body, and tail) and kidneys. Sub-segment masks improve \newloss, but are not necessary (Tab.~\ref{tab:all_results}).}; and the voxel volume $v$ (from CT spacing).

\begin{equation}
\footnotesize
\label{eq:vol_calculation}
    V_{s,o} = \sum_{h,w,l}^{H,W,L} t_{h,w,l} o_{h,w,l} v
\end{equation}

The Volume Loss penalizes the difference between the segmented tumor volume, $V_{s,o}$, and the tumor volume estimated from reports, $V_{r,o}$. To perform this penalization, we tried many losses (e.g., L1 and L2), but Eq.~\ref{eq:custom_loss} converged better:

\begin{equation}
\footnotesize
\label{eq:custom_loss}
L_{\text{forg},o}'(V_{s,o},V_{r,o}) = \frac{|V_{s,o} - V_{r,o}|}{V_{s,o} + V_{r,o} + E}
\end{equation}

The $E$ term increases stability and ensures the loss minimum is at $V_{s,o}=0$ when $V_{r,o}=0$ (we set $E=500$ mm$^3$). Importantly, tumor volumes estimated from reports ($V_{r,o}$) are imperfect. Different radiologists may provide slightly different diameters for a tumor, and we use ellipsoid approximations to convert diameters into volumes. Thus, in Eq.~\ref{eq:vol_loss}, we include a tolerance ($0<\tau<1$; we set $\tau=10\%$) in the Volume Loss: if $|V_{s,o}-V_{r,o}|\lesssim\tau V_{r,o}$, the loss and its gradient become 0, not penalizing the segmenter. The Volume Loss also has a cross-entropy term (Eq.~\ref{eq:volume_bkg}) minimizing tumor segmentations in voxels outside the organ / sub-segment containing tumors ($o$). The final Volume Loss is Eq.~\ref{eq:volume_final}.

{\footnotesize
\begin{gather}
\label{eq:vol_loss}
L_{\text{forg},o}(V_{s,o},V_{r,o}) = \max\{L_{\text{forg},o}'(V_{s,o},V_{r,o})-L_{\text{forg},o}'((1-\tau) V_{r,o},V_{r,o}),0\}\\
\label{eq:volume_bkg}
L_{\text{bkg},o}(\mathbf{T}) = -\frac{1}{H \cdot W \cdot L} \sum_{h=1}^{H} \sum_{w=1}^{W} \sum_{l=1}^{L} \ln(1 - t_{h,w,l}(1-o_{h,w,l})) \\
\label{eq:volume_final}
L_{\text{vol},o} = L_{\text{forg},o}(V_{s,o},V_{r,o}) + L_{\text{bkg},o}(\mathbf{T})
\end{gather}
}

\subsection{Ball Loss}
\label{sec:ball_loss}

\begin{figure}[t]
    \centering
    \includegraphics[width=1\linewidth]{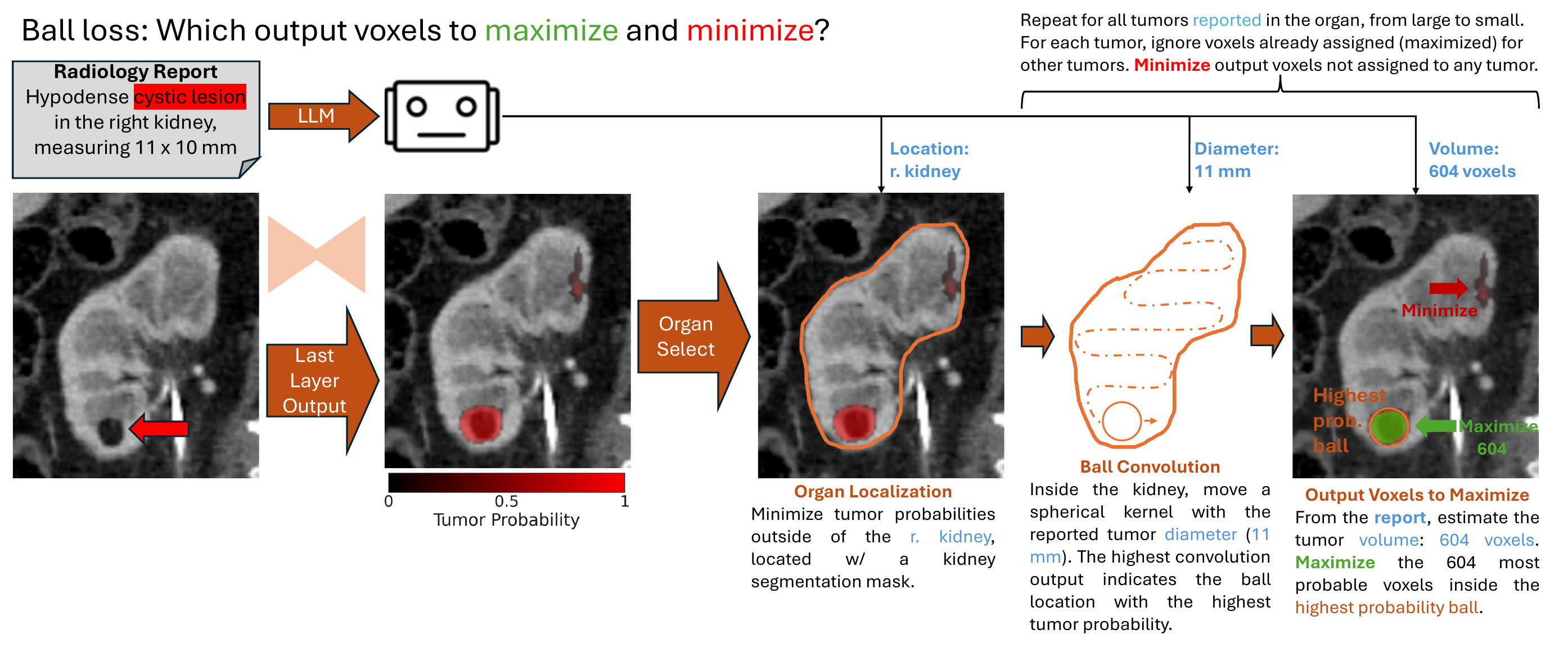}
    \caption{
    \textbf{The Ball Loss enforces segmented tumors to match detailed tumor information from reports---tumor count, locations, diameters, and volumes.} It consists of four steps.
    \textbf{(I)} A large language model (LLM) extracts this tumor information from the report. 
    \textbf{(II)} A Ball Convolution—a simple convolution with a fixed spherical kernel matching the reported tumor diameter—slides over the segmenter output to find the highest-probability tumor region. 
    \textbf{(III)} Within this region, the loss maximizes probabilities in the top-$N$ most probable voxels, where $N$ is the tumor volume estimated from the report (in voxels). These voxels can form any shape that stays within the ball.
    \textbf{(IV)} The process (II--III) is repeated for all reported tumors, largest to smallest. Already-assigned voxels are ignored in later iterations to prevent overlap. 
    Finally, the loss minimizes tumor probabilities in unassigned voxels (background). It focuses more on central tumor voxels and avoids penalizing uncertain tumor borders.
    }
    \label{fig:ball_loss}
\end{figure}

\begin{figure}[t]
    \centering
    \includegraphics[width=\linewidth]{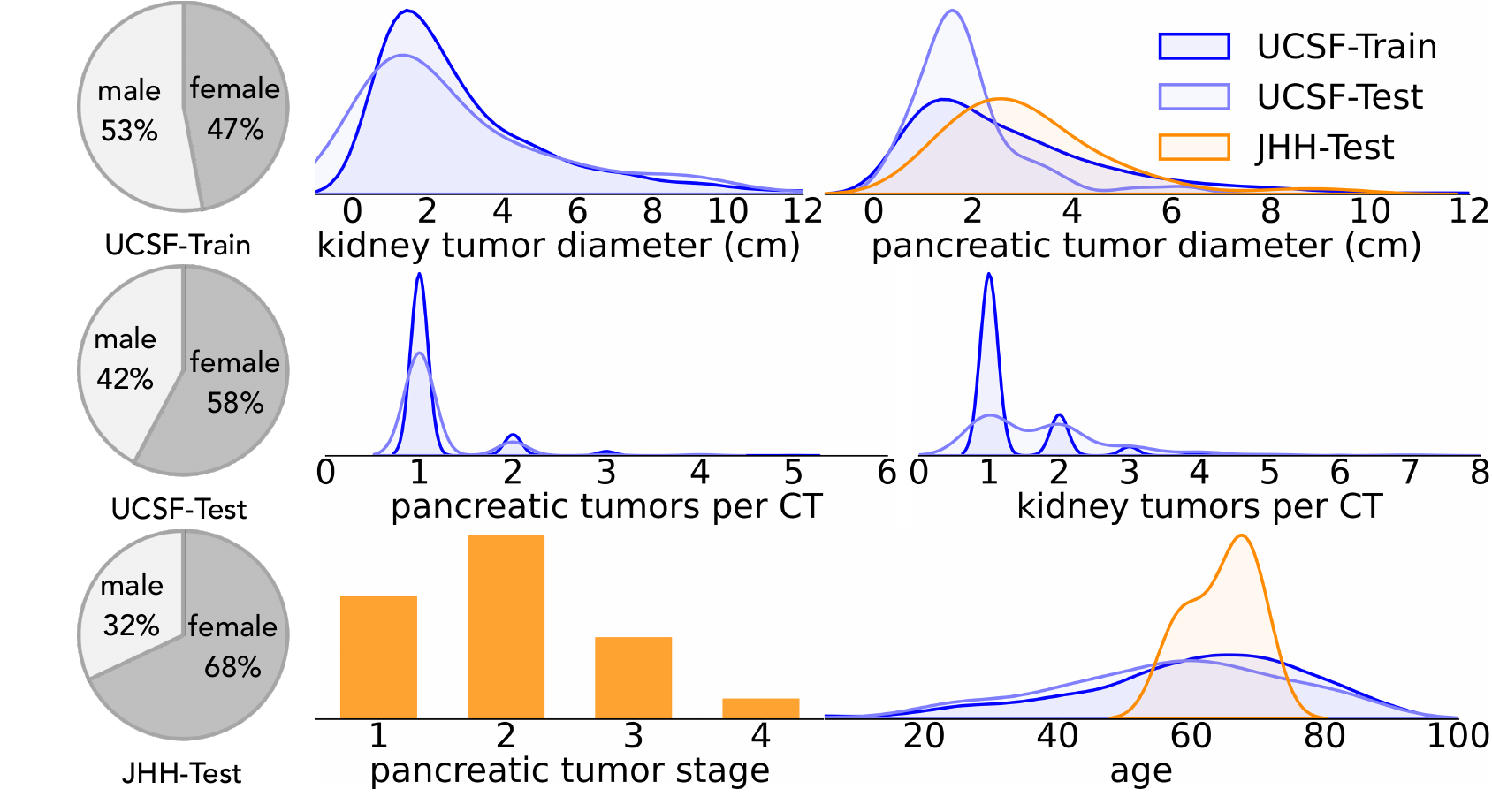}
    \caption{\textbf{\privateOneTrain\ has more pancreatic and kidney tumor CTs than any public CT dataset \cite{bassi2025radgpt}.} We show demographic and tumor data for JHH and UCSF train and test sets (unavailable in AbdomenAtlas 2.0). Reports can be used to easily assemble large datasets \cite{li2025scalemai}, which \method\ directly uses for segmentation training.
    }
    \label{fig:dataset_stats}
\end{figure}
The Ball Loss is applied to the final segmenter output, and enforces the volume, diameter, location, and count of segmented tumors to match reports. The Volume Loss (deep supervision) does not penalize the count or diameter of segmented tumors, allowing the segmenter to explore different possibilities in its intermediate layers. The Ball Loss makes the segmenter refine its prediction. 

Fig.~\ref{fig:ball_loss} shows the Ball Loss. First, we mask the tumor segmentation output ($\mathbf{T}$) with the corresponding organ’s segmentation mask $\mathbf{O}$, giving $\mathbf{T}\otimes\mathbf{O}$. Then, we use Ball Convolutions to locate tumors. This convolution has a non-learnable kernel: a ball matching the tumor diameter in the report (the tumor's largest diameter plus a small margin). For now, assume the kernel is binary: 1 inside the ball, and 0 outside. Convolving the ball over the segmentation output sums tumor probabilities (softmax/sigmoid) within each ball position. Where the sum is highest, we infer the most likely location for a tumor with a diameter similar to the ball. We call the ball in this location the ``\textit{highest probability ball}''. To improve tumor localization, instead of using a binary kernel we weigh the ball kernel higher toward its center, so the convolution responds strongly if the tumor center (where probabilities are usually higher) aligns with the kernel center\footnote{The ball is valued 1 at the center, and decays following a 3D Gaussian with std of $0.75 \times$ the ball diameter, $d_i$. The ball is centered, and the kernel is 0 outside it. For input-output alignment, the convolution has stride 1, zero padding and odd kernels.}.

First, we use the Ball Convolution to locate the largest tumor the report mentions in the organ $o$. The convolution provides the tumor's highest probability ball, and we find the top-$N$ most probable tumor voxels inside it. $N$ is how many voxels the tumor should have, according to the tumor volume estimated from report (\S\ref{sec:volume_loss}). In a blank segmentation mask, we set the top-$N$ voxels to 1. In the segmenter's output, we zero them, to avoid reuse. We repeat the process with the next largest tumor in the report, until all tumors are added to the segmentation mask. The final mask matches the report in tumor count, locations (organs/organ sub-segments), volumes, and diameters. We use these masks as labels for the segmenter, with standard segmentation losses (cross-entropy \& DSC). These dynamic masks improve while the segmenter improves. To compensate for uncertainty and inexact diameters in reports, we give the cross-entropy loss higher weights in tumor voxels with higher predicted tumor probability, and the losses do not penalize a small margin at tumor borders. 

We train with CT patches. If a patch has only part of an organ with a reported tumor, our losses will tell the segmenter to find this tumor, which can be outside the patch. Thus, each of our training patches fully covers one target organ. If a second organ has a reported tumor and is partially inside the patch, we do not apply losses to this organ. If the report mentions a tumor, the ball loss boosts the region with the highest (even if low) predicted tumor probability. When the AI output is near zero everywhere, the volume loss helps: it discourages all-zero outputs with strong gradients (Fig.~\ref{fig:volume_loss}). We did not apply our losses to organs where reports mentioned tumors but not tumor sizes (they were few, <11\%).

\section{Results}\label{sec:results}

\begin{figure}[t]
    \centering
    \includegraphics[width=1\linewidth]{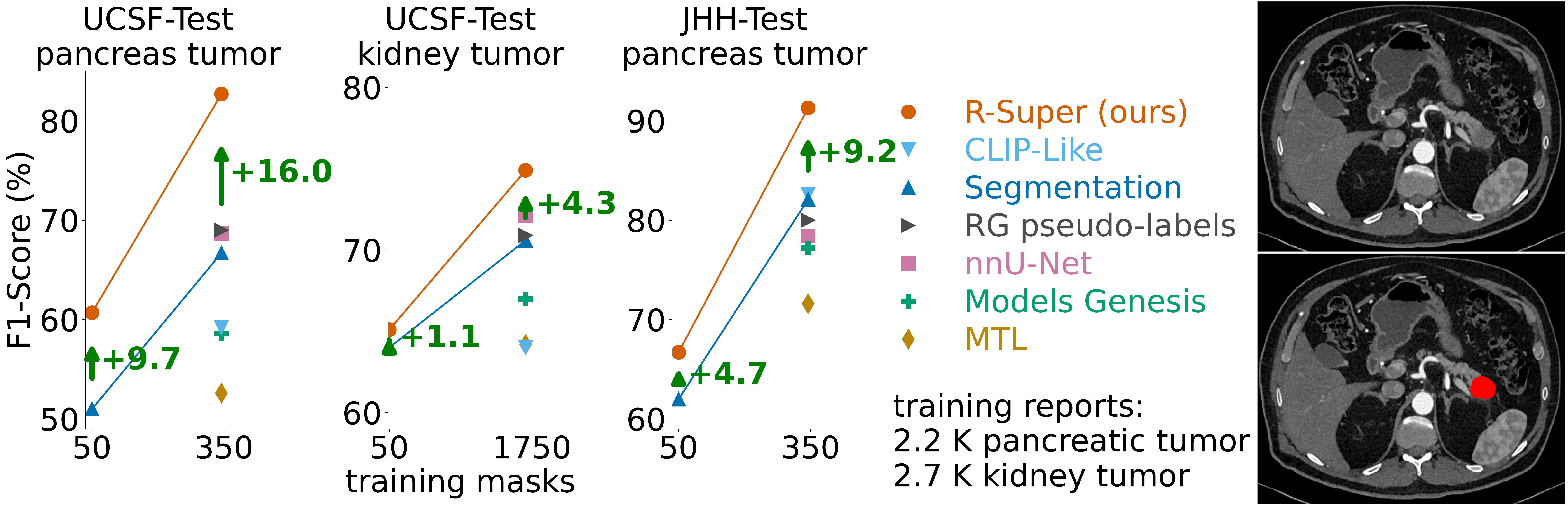}
    \caption{\textbf{By learning from reports, \method\ strongly surpassed 6 state-of-the-art methods in internal (UCSF-Test) and external validation (JHH-Test)}. It surpassed standard segmentation (no report) by up to 16\% in tumor detection F1-Score (green arrows). Each plot corresponds to a tumor type and test set. Their X axes indicate the number of \textit{masks} for training (50 = AbdomenAtlas-Small; 350 / 1.7K = AbdomenAtlas 2.0). \method\ and report-based AIs used masks and 2.2K pancreas/2.7K kidney tumor \textit{reports} for training (\privateOneTrain). Test sets are difficult: JHH-Test is out-of-distribution and UCSF-Test has many CT resolutions and contrast phases (including plain). On the right, there is a pancreatic tumor CT (PDAC) segmented by \method.}
    \label{fig:all_plots}
\end{figure}
\textbf{Datasets.} \textbf{(I) AbdomenAtlas 2.0} \cite{bassi2025radgpt} is one of the largest public CT datasets, containing 9,262 CT-Mask pairs from 88 hospitals across 19 countries, including 344 pancreatic and 1,674 kidney tumor CTs. 
\textbf{(II) AbdomenAtlas-Small} is a subset of AbdomenAtlas 2.0, with 50 pancreatic tumor CTs, 50 kidney tumor CTs, and 100 normal CTs. 
\textbf{(III) \privateOneTrain} is a large private dataset from UCSF and affiliated institutions (California, USA), with \numofct\ CT-Report pairs (no masks): 2,229 with pancreatic tumors, 2,738 with kidney tumors, and 1,751 normals. 
\textbf{(IV) \privateOneTest} is a test set from the same source and distribution as \privateOneTrain, consisting of 169 kidney tumor, 139 pancreatic tumor, and 100 normal CTs.
\textbf{(V) \privateTwo} \cite{xia2022felix,park2020annotated} is an external test set from an unseen hospital—Johns Hopkins Hospital (Maryland, USA)—with 50 pancreatic tumor CTs and 50 normals. Tumors are radiologically confirmed focal lesions in I–IV, and pathology-proven PDACs in V. Datasets I–IV have non-contrast and multiple contrast phases; V contains only arterial phase CT. More details are in Fig.~\ref{fig:dataset_stats}. 

\textbf{Training \method.} We trained two versions of \method. One using \privateOneTrain\ (CT-Report pairs) \& AbdomenAtlas-Small (few CT-Mask pairs), and another using \privateOneTrain\ \& AbdomenAtlas 2.0 (more CT-Mask pairs). In both cases, we first trained on the CT-Mask pairs. Then, we fine-tuned on CT-Mask and CT-Report pairs (\privateOneTrain). We used standard segmentation losses for CT-Mask pairs, and Volume \& Ball losses for CT-Report pairs (setting a weight of 1 to segmentation and 0.1 to our losses, and balancing the number of masks and reports with data augmentation).
We used the MedFormer \cite{gao2022data} segmentation architecture and followed its training hyper-parameters. It is a transformer-CNN hybrid that won previous benchmarks \cite{bassi2024touchstone}; we also used it for all baselines, except nnU-Net \cite{isensee2024nnu}. We tested on \privateOneTest\ (internal) and \privateTwo\ (external). Additionally, we randomly set 10\% of our training set as hold-out validation.

\begin{table}[!t]
\centering
\scriptsize
\caption{\textbf{\method\ surpasses state-of-the-art methods in multiple tumor detection and segmentation metrics.} For pancreatic tumors, \method\ surpasses all alternative methods in DSC, NSD, F1-Score, AUC. For kidney, \method\ matches segmentation training w/ few masks (50), and surpasses it with many (1.7K). DSC and NSD are not available in UCSF-Test, it does not have masks. Sensitivity (Se) and Specificity (Sp) are for thresholds giving highest F1-Scores. The last 2 table rows are ablations: \method\ with only the volume or the ball loss. \method\ and others were trained with pancreas and kidney tumors. Instead, for a pancreas-only \method, Se, Sp, F1, AUC change to 88, 96, 92, 98 in JHH-Test, and 86, 86, 83, 91 in UCSF-Test. Pancreas sub-segment masks improved \method; w/o them (full pancreas masks), F1, AUC drop to 88, 84 in JHH and 74, 80 in UCSF. Orange marks significant gains (p < 0.05, ablations excluded) in F1 (paired permutation test) and AUC (DeLong’s test).}
\begin{tabular}{l*{19}{c}}
\toprule
 & \multicolumn{12}{c}{\footnotesize pancreas tumor} & \multicolumn{6}{c}{\footnotesize kidney tumor} \\
\cmidrule(lr){2-13}\cmidrule(lr){14-19}
 & \multicolumn{8}{c}{\scriptsize JHH-Test} & \multicolumn{4}{c}{\scriptsize UCSF-Test} & \multicolumn{6}{c}{\scriptsize UCSF-Test} \\
\cmidrule(lr){2-9}\cmidrule(lr){10-13}\cmidrule(lr){14-19}
\scriptsize train paradigm & mask & rep. & dsc & nsd & F1 & AUC & Se & Sp & F1 & AUC & Se & Sp & mask & rep. & F1 & AUC & Se & Sp \\
\midrule
\multicolumn{19}{l}{\textit{few training masks (50)}} \\
segmentation~\cite{gao2022data} & 50 & 0 & 38 & 41 & 62 & 63 & 62 & 62 & 51 & 63 & 47 & 77 & 50 & 0 & 64 & 68 & 68 & 63 \\
\textbf{R-Super (ours)} & 50 & 2.2K & 49 & 52 & 67 & 75 & 68 & 64 & 61 & 70 & 73 & 59 & 50 & 2.7K & 65 & 66 & 73 & 55 \\
\midrule
\multicolumn{19}{l}{\textit{medium / many training masks (344 / 1.7K)}} \\
CLIP-Like \cite{blankemeier2024merlin} & 344 & 2.2K & 55 & 58 & 83 & 86 & 90 & 70 & 59 & 74 & 59 & 75 & 1.7K & 2.7K & 64 & 71 & 75 & 48 \\
Multi-task l. \cite{chen2019lesion} & 344 & 2.2K & 38 & 43 & 72 & 80 & 78 & 60 & 53 & 62 & 65 & 50 & 1.7K & 2.7K & 64 & 71 & 68 & 63 \\
RG Pseudo-l. \cite{bosma2023semisupervised} & 344 & 2.2K & 56 & 62 & 80 & 79 & 80 & 78 & 69 & 83 & 65 & 86 & 1.7K & 2.7K & 71 & 74 & 80 & 60 \\
Models G. \cite{zhou2021models} & 344 & 0 & 53 & 57 & 77 & 81 & 78 & 74 & 59 & 72 & 66 & 63 & 1.7K & 0 & 67 & 73 & 77 & 54 \\
nnU-Net \cite{isensee2021nnu} & 344 & 0 & 53 & 57 & 78 & 75 & 76 & 82 & 69 & 79 & 74 & 74 & 1.7K & 0 & 72 & 74 & 87 & 53 \\
segmentation~\cite{gao2022data} & 344 & 0 & 51 & 59 & 82 & 84 & 78 & 88 & 67 & 78 & 62 & 85 & 1.7K & 0 & 71 & 73 & 65 & 68 \\
\textbf{R-Super (ours)} & 344 & 2.2K & 59 & 69 & \cellcolor{lightorange}91 & \cellcolor{lightorange}92 & 94 & 88 & \cellcolor{lightorange}83 & \cellcolor{lightorange}90 & 83 & 89 & 1.7K & 2.7K & \cellcolor{lightorange}75 & \cellcolor{lightorange}78 & 80 & 70 \\
\midrule
\multicolumn{19}{l}{\textit{ablation studies}} \\
Vol.~Loss (ours) & 344 & 2.2K & 59 & 61 & 94 & 97 & 90 & 98 & 76 & 89 & 80 & 82 & 1.7K & 2.7K & 75 & 74 & 86 & 63 \\
Ball Loss (ours) & 344 & 2.2K & 59 & 59 & 83 & 92 & 78 & 90 & 71 & 82 & 64 & 90 & 1.7K & 2.7K & 74 & 75 & 87 & 58 \\
\bottomrule
\end{tabular}
\label{tab:all_results}
\end{table}


\begin{table}[!t]
\centering
\scriptsize
\caption{\textbf{\method\ surpasses state-of-the-art methods in detecting both small (diameter $\leq$ 2 cm) and large tumors.} For small and large pancreatic tumors, \method\ surpasses all alternative methods in DSC, NSD, F1-Score, and AUC (344 training masks). For small kidney tumors, \method\ surpasses the alternative methods in F1-Score and AUC, and in F1-Score for large tumors (1.7K training masks). DSC and NSD are not available in UCSF-Test (no masks). Sensitivity (Se) and Specificity (Sp) are given for thresholds giving highest F1-Scores. The last 2 table rows are ablations: \method\ with only the volume or the ball loss. Orange marks significant gains (p < 0.05, ablations excluded) in F1-Score (paired permutation test) and AUC (DeLong’s test). Test set sizes: JHH-Test has 15 small and 35 large pancreatic tumors; and UCSF-Test has 90 small and 28 large pancreatic tumors (21 CTs with tumors of unknown size were excluded from this table), and 77 small and 65 large kidney tumors (27 excluded).}
\begin{tabular}{l*{19}{c}}
\toprule
 & \multicolumn{12}{c}{\footnotesize pancreas tumor} & \multicolumn{6}{c}{\footnotesize kidney tumor} \\
\cmidrule(lr){2-13}\cmidrule(lr){14-19}
 & \multicolumn{8}{c}{\scriptsize JHH-Test} & \multicolumn{4}{c}{\scriptsize UCSF-Test} & \multicolumn{6}{c}{\scriptsize UCSF-Test} \\
\cmidrule(lr){2-9}\cmidrule(lr){10-13}\cmidrule(lr){14-19}
\scriptsize train paradigm & mask & rep. & dsc & nsd & F1 & AUC & Se & Sp & F1 & AUC & Se & Sp & mask & rep. & F1 & AUC & Se & Sp \\
\midrule
\multicolumn{19}{c}{\textbf{small tumors (diameter $\leq$ 2 cm)}} \\
\midrule
\addlinespace[2pt]
\multicolumn{19}{l}{\textit{few training masks (50)}} \\
segmentation~\cite{gao2022data} & 50 & 0 & 7 & 17 & 38 & 61 & 53 & 62 & 41 & 59 & 40 & 77 & 50 & 0 & 43 & 59 & 53 & 63 \\
\textbf{R-Super (our)} & 50 & 2.2K & 15 & 25 & 35 & 68 & 47 & 64 & 54 & 72 & 76 & 59 & 50 & 2.7K & 47 & 61 & 65 & 55 \\
\midrule
\multicolumn{19}{l}{\textit{medium / many training masks (344 / 1.7K)}} \\
CLIP-Like \cite{blankemeier2024merlin} & 344 & 2.2K & 11 & 19 & 68 & 90 & 100 & 71 & 50 & 74 & 54 & 75 & 1.7K & 2.7K & 40 & 65 & 57 & 48 \\
Multi-task l. \cite{chen2019lesion} & 344 & 2.2K & 15 & 26 & 54 & 83 & 87 & 60 & 42 & 61 & 60 & 50 & 1.7K & 2.7K & 46 & 65 & 57 & 63 \\
RG Pseudo-l. \cite{bosma2023semisupervised} & 344 & 2.2K & 19 & 32 & 61 & 77 & 73 & 80 & 63 & 82 & 62 & 86 & 1.7K & 2.7K & 50 & 71 & 66 & 60 \\
Models G. \cite{zhou2021models} & 344 & 0 & 10 & 20 & 62 & 85 & 80 & 76 & 48 & 70 & 60 & 63 & 1.7K & 0 & 43 & 65 & 60 & 54 \\
nnU-Net \cite{isensee2021nnu} & 344 & 0 & 7 & 17 & 55 & 74 & 60 & 82 & 60 & 78 & 70 & 74 & 1.7K & 0 & 53 & 71 & 79 & 53 \\
segmentation~\cite{gao2022data} & 344 & 0 & 17 & 36 & 65 & 83 & 67 & 88 & 59 & 77 & 58 & 85 & 1.7K & 0 & 39 & 69 & 44 & 68 \\
\textbf{R-Super (our)} & 344 & 2.2K & 25 & 48 & 80 & \cellcolor{lightorange}89 & 93 & 88 & \cellcolor{lightorange}75 & \cellcolor{lightorange}90 & 77 & 89 & 1.7K & 2.7K & 56 & \cellcolor{lightorange}78 & 69 & 69 \\
\midrule
\multicolumn{19}{l}{\textit{ablation studies}} \\
Vol.~Loss (our) & 344 & 2.2K & 27 & 32 & 93 & 99 & 93 & 98 & 69 & 87 & 77 & 82 & 1.7K & 2.7K & 57 & 73 & 77 & 63 \\
Ball Loss (our) & 344 & 2.2K & 20 & 22 & 67 & 89 & 67 & 90 & 64 & 78 & 60 & 90 & 1.7K & 2.7K & 56 & 71 & 80 & 58 \\
\midrule[1pt]
\multicolumn{19}{c}{\textbf{large tumors (diameter > 2 cm)}} \\
\midrule
\addlinespace[2pt]
\multicolumn{19}{l}{\textit{few training masks (50)}} \\
segmentation~\cite{gao2022data} & 50 & 0 & 3 & 5 & 59 & 63 & 65 & 62 & 35 & 78 & 61 & 77 & 50 & 0 & 56 & 75 & 82 & 63 \\
\textbf{R-Super (our)} & 50 & 2.2K & 8 & 13 & 67 & 78 & 76 & 64 & 31 & 75 & 79 & 59 & 50 & 2.7K & 54 & 72 & 86 & 55 \\
\midrule
\multicolumn{19}{l}{\textit{medium / many training masks (344 / 1.7K)}} \\
CLIP-Like \cite{blankemeier2024merlin} & 344 & 2.2K & 13 & 19 & 75 & 84 & 85 & 71 & 43 & 84 & 82 & 75 & 1.7K & 2.7K & 54 & 77 & 95 & 48 \\
Multi-task l. \cite{chen2019lesion} & 344 & 2.2K & 18 & 28 & 63 & 78 & 74 & 60 & 28 & 70 & 82 & 50 & 1.7K & 2.7K & 54 & 77 & 78 & 63 \\
RG Pseudo-l. \cite{bosma2023semisupervised} & 344 & 2.2K & 19 & 31 & 78 & 79 & 82 & 80 & 55 & 90 & 82 & 86 & 1.7K & 2.7K & 62 & 75 & 97 & 60 \\
Models G. \cite{zhou2021models} & 344 & 0 & 13 & 21 & 72 & 79 & 76 & 76 & 36 & 83 & 86 & 63 & 1.7K & 0 & 58 & \cellcolor{lightorange}80 & 97 & 54 \\
nnU-Net \cite{isensee2021nnu} & 344 & 0 & 11 & 19 & 79 & 75 & 82 & 82 & 46 & 89 & 93 & 74 & 1.7K & 0 & 57 & 76 & 97 & 53 \\
segmentation~\cite{gao2022data} & 344 & 0 & 18 & 33 & 82 & 84 & 82 & 88 & 56 & 88 & 86 & 85 & 1.7K & 0 & 63 & 75 & 91 & 68 \\
\textbf{R-Super (our)} & 344 & 2.2K & 33 & 51 & 89 & \cellcolor{lightorange}93 & 94 & 88 & \cellcolor{lightorange}68 & \cellcolor{lightorange}94 & 96 & 89 & 1.7K & 2.7K & \cellcolor{lightorange}67 & 75 & 97 & 69 \\
\midrule
\multicolumn{19}{l}{\textit{ablation studies}} \\
Vol.~Loss (our) & 344 & 2.2K & 35 & 37 & 92 & 97 & 88 & 98 & 54 & 96 & 93 & 82 & 1.7K & 2.7K & 63 & 74 & 97 & 63 \\
Ball Loss (our) & 344 & 2.2K & 25 & 25 & 85 & 93 & 85 & 90 & 62 & 94 & 82 & 90 & 1.7K & 2.7K & 61 & 78 & 98 & 58 \\
\bottomrule
\end{tabular}
\label{tab:results_by_size}
\end{table}

\textbf{\method\ improves tumor segmentation with few and many training masks.} Fig.~\ref{fig:all_plots} compares \method\ to standard segmentation (trained w/o reports). By including 2.2K pancreatic and 2.7K kidney tumor CT-Report pairs in training, \method\ strongly surpassed segmentation when the training set had few (50 in AbdomenAtlas-Small) training masks (up to +9.7\% F1-Score), medium (344 in AbdomenAtlas 2.0) masks (up to +16\% F1-Score), and many (1.7K in AbdomenAtlas 2.0) masks (+4.3\% F1-Score). Thus, \method\ can use radiology reports to help segmenting tumor types with scarce segmentation masks, and it can scale up already large tumor segmentation datasets. 

\textbf{\method\ improves tumor segmentation on the hospital that provided training reports and on an unseen hospital.} Fig.~\ref{fig:all_plots} shows that \method\ strongly surpasses segmentation both on \privateOneTest\ (same hospital as \privateOneTrain), and on \privateTwo, a dataset from a hospital never seen during training. Interestingly, results on \privateTwo\ surpassed \privateOne\ (Fig.~\ref{fig:all_plots}), possibly due to finer CT slice thickness (0.5 mm vs. many thickness in \privateOneTest), and arterial contrast in \privateTwo\ (vs. many contrast phases \& non-contrast in \privateOne).  In both test datasets, \method\ strongly surpasses six state-of-the-art methodologies, including 4 that also use reports in training (Tab.~\ref{tab:sota_comparison}).
\method\ was superior not only in tumor detection F1-Score (Fig.~\ref{fig:all_plots}), but also in AUC, DSC and NSD (Tab.~\ref{tab:all_results}). Furthermore, \method\ strongly improved AI performance for both small tumors---diameter $\leq$2 cm, important for early cancer detection---and large tumors (Tab. \ref{tab:results_by_size}). Thus, \method's advantages over the alternative methods (explained in Tab.~\ref{tab:sota_comparison}) translated into overall superior tumor segmentation.
\section{Conclusion}\label{sec:conclusion}

\method\ transforms radiology reports into supervision for segmentation, allowing us to scale segmentation AI by leveraging the vast amount of CT–Report pairs in hospitals and upcoming public datasets \cite{hamamci2024ct2rep,bassi2025radgpt}---when few or many masks are available. It also allows easier merging of data from many hospitals, enhancing data diversity and AI generalization. Here, merging the \privateOneTrain\ CT-Report dataset with AbdomenAtlas 2.0 yielded a training set with 2,573 pancreatic and 4,412 kidney tumor CTs---way more than any public set. Yet, a single hospital can have orders of magnitude more reports. Therefore, the exceptional \method\ results in this study (e.g., +16\% F1-Score in Fig.~\ref{fig:all_plots}) are an initial sample---and an unveiling---of the transformative potential of report-driven segmentation AI. 

\begin{credits}
\subsubsection{\ackname} This work was supported by the Lustgarten Foundation for Pancreatic Cancer Research, the Patrick J. McGovern Foundation Award, and the National Institutes of Health (NIH) under Award Number R01EB037669. We would like to thank the Johns Hopkins Research IT team in \href{https://researchit.jhu.edu/}{IT@JH} for their support and infrastructure resources where some of these analyses were conducted; especially \href{https://researchit.jhu.edu/research-hpc/}{DISCOVERY HPC}. P.R.A.S.B. thanks the funding from the Center for Biomolecular Nanotechnologies, Istituto Italiano di Tecnologia (73010, Arnesano, LE, Italy). No interests to disclose.
\end{credits}

\clearpage
\bibliographystyle{splncs04}
\bibliography{refs,zzhou}

\end{document}

%% file: preamble.tex
\usepackage{float}
\usepackage{pifont}
\usepackage{footnote}
\usepackage{enumitem}
\usepackage{bm}
\usepackage{arydshln}
\usepackage{booktabs}
\usepackage{multicol}
\usepackage{multirow}
\usepackage{color}
\usepackage{xcolor}     
\usepackage{colortbl}
\usepackage{soul}
\usepackage{bbding}
\usepackage{makecell}
\usepackage{mathtools}
\usepackage{imakeidx}
\usepackage{amssymb}
\usepackage{graphicx}
\usepackage{amsmath}
\usepackage{threeparttable}
\definecolor{citecolor}{HTML}{0071BC}
\definecolor{linkcolor}{HTML}{ED1C24}
\usepackage[colorlinks,
            anchorcolor=red,
            citecolor=citecolor, 
            linkcolor=linkcolor,
            ]{hyperref}
\makeindex
\usepackage{arydshln}
\usepackage{lipsum}
\usepackage[toc]{multitoc}
\usepackage[edges]{forest}
\usepackage[normalem]{ulem}

\usepackage{bbding}
\usepackage[most]{tcolorbox}

\usepackage{algorithm}
\usepackage{algorithmic}

\usepackage{minitoc}
\usepackage[toc,page,header]{appendix}

\definecolor{orchid}{rgb}{0.85, 0.44, 0.84}
\definecolor{rubinred}{rgb}{0.82, 0.0, 0.28}
\definecolor{flagship}{rgb}{0.93, 0.06, 0.41}
\definecolor{radiologist}{rgb}{0.50, 0.50, 1}

\definecolor{YT}{HTML}{002FA7}

\newcommand{\method}{{\fontfamily{ppl}\selectfont
R-Super}}
\newcommand{\newloss}{{\fontfamily{ppl}\selectfont
R-Super}}

\newcommand{\privateOne}{UCSF-Set}
\newcommand{\privateOneTest}{UCSF-Test}
\newcommand{\privateOneTrain}{UCSF-Train}
\newcommand{\privateTwo}{JHH-Test}

\newcommand{\numofct}{6,718}

\newcolumntype{P}[1]{>{\centering\arraybackslash}p{#1}}
\newlength\savewidth

\newcommand{\cmark}{\textcolor{BurntOrange}{\ding{51}}} %
\definecolor{lightgray}{gray}{0.8} 
\newcommand{\xmark}{\textcolor{lightgray}{\ding{55}}}   %